\documentstyle[12pt,aaspp4]{article}

\begin{document}

   \title{Cepheid pulsation models at varying metallicity and  $\Delta{Y}/\Delta{Z}$}

\author{M. Marconi \altaffilmark{1}, I. Musella \altaffilmark{1},
G. Fiorentino \altaffilmark{2,3}}

\affil{1. Osservatorio Astronomico Di Capodimonte, Via Moiariello 16,
80131 Napoli, Italy; marcella@na.astro.it; ilaria@na.astro.it}

\affil{2. Osservatorio Astronomico di Roma, Via Frascati 33, 00040 Monte Porzio Catone, Italy; giuliana@mporzio.astro.it}

\affil{3. Universit\`a degli Studi di Roma ``Tor Vergata'', Via 
Della Ricerca Scientifica 1, 00133 Roma, Italy}

\begin{abstract}   
In this paper we present an extended set of nonlinear convective
pulsation models at varying the metallicity and $\Delta{Y}/\Delta{Z}$
ratio. The predicted instability strip and bolometric light curves are
discussed by comparing the new models with our previous ones. In
particular, the dependence on both metal and helium abundances is
investigated. By transforming the bolometric light curves into the
observational bands we are able to derive both Period-Color-Luminosity
and Wesenheit relations for each selected chemical
composition. Synthetic Period-Luminosity relations are obtained by
populating the instability strip according to specific assumptions on
the number of pulsators and the mass distribution. These theoretical
results are compared with recent accurate data by Sandage et al. and 
Kervella et al., in order to test the predictive capabilities of the
models. We confirm our previous results that the theoretical
metallicity correction to the Key Project Cepheid distance scale
depends both on the period range and $\Delta{Y}/\Delta{Z}$ ratio,
becoming important for periods longer than 20 days and
$\Delta{Y}/\Delta{Z} > 1.5$.


\end{abstract}   


\section{Introduction}   

Classical Cepheids are the most reliable primary distance indicators
for Local Group and (from the space) external galaxies, thanks to
their characteristic period-luminosity-color (PLC) and
period-luminosity (PL) relations. Moreover, through the calibration of
secondary distance indicators, they allow us to reach cosmological
distances (of the order of 100 Mpc), thus providing fundamental
constraints on the Hubble constant (see e.g. Freedman et al. 2001,
hereinafter F01; Saha et al. 2001). The problem of the dependence of
the Cepheid PL relation on chemical composition has been widely
debated in the recent literature but with quite different results,
depending on the adopted method and the authors (see e.g. Kennicutt et
al. 1998, Fiorentino et al. 2002, Storm et al. 2004, Groenewegen et
al. 2004, Sakai et al. 2004, Romaniello et al. 2005).  In the last few
years we have studied the Cepheid pulsation properties through the
computation of nonlinear, nonlocal and time-dependent convective
pulsation models, which allow to predict all the relevant pulsational
observables. In particular, the nonlinearity and the inclusion of a
detailed treatment of the coupling between pulsation and convection
allow these models to predict not only the periods and the blue
boundary of the instability strip, but also the pulsation amplitudes,
the detailed light and radial velocity curve morphology and the
complete topology of the strip, including the red edge which is caused
by the pulsation quenching due to convection (see Bono, Marconi \&
Stellingwerf 1999a, 2000a and references therein for details). On this
basis, various sets of Cepheid models have been computed with varying
chemical composition ($0.004<$Z$<0.04$, $0.25<$Y$<0.33$) and stellar
mass from 2.8 M$_{\odot}$ to 11 M$_{\odot}$ (Bono et al. 1999b, 2000b,
2002b; Fiorentino et al. 2002, hereinafter F02). For each chemical
composition and mass, an evolutionary mass-luminosity (ML) relation
was adopted (see F02 for details) and a wide range of effective
temperature was explored. As a result, we have found that the Cepheid
properties, and in particular the location in the HR diagram of the
instability strip and the coefficients of the multiband PL relations,
depend on the pulsator metallicity with the amplitude of the effect
decreasing from visual to near infrared magnitudes (Caputo, Marconi \&
Musella, 2000a, hereinafter C00). In particular, as the model
metallicity increases from $Z=0.004$ to $Z=0.03$ the instability strip
gets redder and the pulsator luminosity, at fixed period, gets fainter
(Bono et al. 1999b, Caputo et al. 2000b, F02). This result is at
variance with recent empirical evaluations of the metallicity effect
(see e.g. Kennicutt et al. 1998; F01) and relies on the assumption of
$\Delta{Y}/\Delta{Z}=2.5$. Specific computations for
$\Delta{Y}/\Delta{Z}=4.0$ have shown that, at least for the higher
metal contents ($Z\ge{0.008}$), the location into the HR diagram of
the Cepheid instability strip also depends on helium abundance, moving
toward higher effective temperature as Y increases, at fixed Z (F02).
On the basis of the above chemical composition dependent models, F02
have found that the adoption of LMC based $V$ and $I$ PL relations to
get distance moduli with an uncertainty of $\pm 0.1$ mag is justified
for variables with period shorter than 10 days. At longer periods, a
correction to LMC based distances maybe needed, whose sign and amount
depend on the helium and metal content of the Cepheids. In particular,
model predictions were found to account for the empirical metallicity
correction suggested by Kennicutt et al. (1998), provided that the
adopted helium-to-metal enrichment ratio was about 3.5.  Moreover, the
above models provide a fairly good description of the data obtained by
Romaniello et al. (2005, hereinafter R05), by relating the $V$ band
residuals from the PL relation adopted by the HST Key Project (F01) to
spectroscopic iron abundances measured for 37 Galactic and Magellanic
Clouds Cepheids (see R05 for details).  All the above results were
essentially based on 2 values (at most) of helium content for each
fixed metal abundance and therefore did not allow to properly
investigate the helium effect on the whole metal content range of
observed Cepheids. In order to perform a more accurate analysis of
combined helium and metal effects, we extended our grid of models to
other chemical compositions at varying $\Delta Y/\Delta Z$.  In this
paper we show the results of these new computations and further
discuss the effect of chemical composition on Cepheid properties. In
Sect. 2 we present the new model set and combine the results with the
previous ones to investigate the dependence of the predicted pulsation
observables on helium abundance and metallicity. In Sect. 3 the
theoretical PL, PLC and Wesenheit relations are presented, whereas in
Sect. 4 and 5 we discuss the comparison with recent observational data
and the implication of the predictions presented in the previous
sections for the Cepheid distance scale.

\section{The new models}   

By adopting the same code and physical and numerical assumptions as in
previous papers (Bono et al. 1999a, F02), we have computed new
sequences of pulsation models for the input parameters listed in Table
1. In particular these new models correspond to different values of
the $\Delta Y/\Delta Z$ parameter, ranging from 0.5 to 3.5. The upper
limit is due to the evidence that, as noticed in F02, for $\Delta
Y/\Delta Z = 4$ no model is found to pulsate at the highest
metallicities ($Z\sim0.04$) covered by Cepheids in HST galaxies.  On
the other hand, $\Delta Y/\Delta Z = 0.5$ is below the lower limit of
the evaluations found in the recent literature (see e.g. Pagel \&
Portinari 1998, Pagel et al. 1992, Izotov, Thuan, \& Lipovetsky 1997,
Izotov \& Thuan 2004).  For each selected mass, the luminosity level
is chosen on the basis of the same canonical evolutionary ML relation
adopted in F02. For a detailed discussion of  the effect of the ML
selection on Cepheid properties, we refer the interested reader to a
companion paper by Caputo et  al. (2005, hereinafter C05).  A wide
range of effective temperatures  is explored for each model mass and
the modal stability is  investigated for the fundamental mode. The
first overtone mode  pulsation is not studied in this paper because it
is not expected to  be significant at the selected metallicity and
mass ranges and because  it is known to be almost independent on
chemical composition (see Bono  et al. 2001, 2002b and references
therein for details).   

\begin{table} 
\caption{Input parameters of the new  pulsation models.} 
\begin{center} 
\vspace{0.5truecm} 
\begin{tabular}{llccc} 
\hline \hline $Y$ & $Z$ & $M/M_{\odot}$ & log$L/L_{\odot}$ & $\Delta Y/\Delta Z$ \\ \hline 0.26 & 0.01 & 5.00& 3.13 &  2.5 \\ " 
      & " & 7.00 & 3.61  &"  \\ " & " & 9.00 & 3.98  &"  \\ " & " &
      11.00 & 4.27  &"  \\ 0.25 & 0.02 & 5.00& 3.00 &  1 \\ " &
      " & 7.00 & 3.49  & "  \\ " & " & 9.00 & 3.86  & "  \\ " & " &
      11.00 & 4.15  & "  \\ 0.26 & 0.02 & 5.00& 3.02 & 1.5 \\ "
      & "& 7.00 & 3.51  & "  \\ " & "& 9.00 & 3.88 & "  \\ " & " &
      11.00 & 4.17 & "  \\ 0.275 & 0.03 & 5.00& 3.00 & 1.5 \\ "
      & "& 7.00 & 3.49  & "  \\ " & "& 9.00 & 3.85 & "  \\ " & " &
      11.00 & 4.14 & "  \\ 0.335 & 0.03 & 5.00& 3.11 &  3.5 \\ "
      & "& 7.00 & 3.60  & "  \\ " & "& 9.00 & 3.97 & "  \\ " & " &
      11.00 & 4.26 & "  \\ 0.25 & 0.04 & 5.00& 3.06 &  0.5 \\ "
      & "& 7.00 & 3.55  & "  \\ " & "& 9.00 & 3.92 & "  \\ " & " &
      11.00 & 4.21 & "  \\ 0.29 & 0.04 & 5.00& 2.99 &  1.5 \\ "
      & "& 7.00 & 3.47  & "  \\ " & "& 9.00 & 3.84 & "  \\ " & " &
      11.00 & 4.13 & "  \\ 
\hline
\end{tabular} 
\end{center} 
\end{table}

\subsection{The instability strip}   

For each chemical composition, mass and luminosity level, model
computations allowed us to derive the blue (FBE) and red (FRE) edges
of the fundamental instability strip. The new strips are reported in
Figs. 1 and 2 together with similar evaluations from our previous
model sets.  In Fig. 1 we show the location of the predicted
instability strip in the HR diagram at fixed $\Delta{Y}/\Delta{Z}$ and
for the labeled metal contents. This plot confirms our previous
results (see F02 and references therein) concerning the shift of the
instability strip toward lower effective temperatures, as the metal
abundance increases from $Z=0.004$ to $Z=0.03$, and the narrowing of the
strip when passing from $Z=0.03$ to $Z=0.04$. The latter occurrence is due
to the reduced efficiency of pulsation associated to the low Hydrogen
abundance for $Z=0.04$ and $Y=0.29$ or 0.33 (corresponding to
$\Delta{Y}/\Delta{Z}=1.5$ or 2.5 respectively).  In order to
investigate the effect of varying the helium abundance at fixed
metallicity, in the three panels of Fig. 2 we show the location of the
instability strip in the HR diagram for $Z=0.02$ (top), $Z=0.03$ (middle)
and $Z=0.04$ (bottom) and the labeled helium abundances. For $Z=0.04$ we
note a narrowing of the instability strip when Y increases (and X
decreases). The effect is less evident for the other two metal
abundances. In particular, for $Z=0.02$ the FRE moves toward higher
effective temperature as Y increases from $Y=0.25$ to $Y=0.31$, confirming
the result found by F02, but the helium dependence of the FBE is much
more complicated with a sort of to and fro behavior on the explored Y
range. This occurrence is related to the competing pulsation driving
role of H and He abundances in the associated ionization zones (see
Bono et al. 1999a for details).

\subsection{The light and radial velocity curves}   

One of the most important output of nonlinear pulsation codes is the
predicted variation of relevant quantities (luminosity, radial
velocity, effective temperature, surface gravity) along a model
pulsation cycle.  The bolometric light curves\footnote{The radial
velocity curves are also available upon request to the authors.} for
the models quoted in Table 1 are reported in Figs 3a-3g for each
labeled mass and luminosity level. The model period and effective
temperature is also reported in each panel.  We notice that both the
morphology and the amplitude of the curves vary with the position
within the instability strip and depend on the adopted chemical
composition.  This behavior confirms our previous results (see Bono,
Castellani \& Marconi 2000b, Bono, Marconi \& Stellingwerf 2000a,
hereinafter BMS00). In particular, these plots support the empirical
evidence for Galactic Cepheids originally found by Sandage
\& Tammann (1968, 1971) and Cogan (1980), and recently confirmed on
the basis of a much larger sample of Cepheids by Sandage, Tammann \&
Reindl (2004) and Tammann, Sandage \& Reindl (2003), that, in the
period range $\log{P}\approx{0.40-0.86}$ and for $\log{P}>1.1-1.3$ the
largest luminosity amplitudes are attained close to the blue edge,
while for $0.85 < \log{P} < 1.1-1.3 $, the maximum is attained close to
the red edge as a consequence of a phenomenon called Hertzsprung
progression (HP) (Hertzsprung 1926; Ledoux \& Walraven
1958). Classical Cepheids in the period range $6 <P<16$~d show a
secondary maximum (bump) along both the light and the radial velocity
curves. The HP is the relationship between the phase of this bump and
the pulsation period. In particular, for Galactic Cepheids the bump
appears on the descending branch of the light curve for Cepheids with
periods up to $\sim 9$~d, while it appears close to maximum light for
$9 <P<12$~d and moves at earlier phases for longer periods. On the
basis of this observational evidence this group of variables was
christened ``Bump Cepheids''. As already obtained by BMS00 for
$0.004<Z<0.02$, inspection of Figs. 3a, 3b, 3c and 3f suggest that an
increase in the metal content causes a shift of the HP center toward
shorter periods. In fact, as shown more in detail in Fig. 4, passing
from $Z=0.01$ $Y=0.26$ (upper panel) to $Z=0.04$ $Y=0.25$ (bottom
panel), for $M/M\odot =7$, the period corresponding to the HP center
moves from $\sim 10.5$~d to $\sim 8.2$~d, attaining $\sim 9.5$~d for
$Z=0.02$ $Y=0.25$,0.26 (middle panels). We remind that this trend is
in agreement with the observations. In fact, empirical data for
Galactic Cepheids suggest that the HP center corresponds to a period
PHP$\sim 10.0$~d (Moskalik, Buchler \& Marom 1992; Moskalik et
al. 2000), whereas in the LMC ($Z=0.008$) PHP$\sim 10.5$~d and in the
SMC ($Z=0.004$) PHP$\sim 11.0$~d (Beaulieu 1998). Unfortunately, no
empirical evidence is available for the HP phenomenon in sovrasolar
Cepheid samples.  The bolometric light curves of the new models were
transformed into the observational bands ($UBVRIJK$) by means of the
model atmospheres by Castelli, Gratton \& Kurucz (1997a,b) and mean
magnitudes and colors were then derived for each chemical composition
and stellar mass. In Table 2 we report the model periods and intensity
weighted mean magnitudes, but magnitude averaged values are also
available upon request to the authors. The static magnitude
values\footnote{For static magnitude we mean the magnitude the star
would have were it not pulsating} have also been derived and used to
obtain the boundaries of the instability strip, at each chemical
composition, in the various period-magnitude planes and, in turn, to
construct synthetic multiband PL relations (see below).
\section{Predicted Cepheid relations} The results presented in the
previous section allow us to derive all the relevant relations
connecting the pulsation period to mean magnitudes and colors, as well
as synthetic PL relations, following the same procedure as in our
previous papers (see Caputo et al. 2000a, hereinafter C00,F02).

\subsection{PLC and Wesenheit relations}   

Linear regression through the period and magnitude values reported in
Table 2 provides the multiband PLC relations given in Table 3. In the
same table we also report the PLC coefficients of our previous model
sets.  Similarly, the coefficients of the reddening free Wesenheit
relations (see C00 and references therein) are reported in Table
4. These are defined by using the ratios between total
extinction and the various color excesses given by Cardelli et
al. (1989).  In agreement with the recent empirical evidence by Ngeow
\& Kanbur (2005), we find that the predicted Wesenheit relations are
well represented by linear functions, at variance with PL (see below)
and Period-Color (see C00) relations. We notice that, even if PLC and
Wesenheit relations have the advantage of being independent on the
distribution of pulsators within the instability strip, holding for
each individual star as a result of its period-density relation and
black body behavior, they heavily rely on the assumption of an
evolutionary ML relation.  Without this assumption we would obtain
tight mass-dependent PLC and Wesenheit relations (see C05) which allow
to provide sound constraints on the pulsation mass of each individual
Cepheid, once known the absolute magnitudes and the intrinsic colors,
and, by comparison with evolutionary masses, to give an estimate of
mass-loss during or before the Cepheid phase. The interested reader is
referred to C05 for a detailed and updated investigation of this
problem.  Here we only notice that if we used a non-canonical
mass-luminosity relation\footnote{By non-canonical we mean based on
evolutionary models including a mild core overshooting during the
hydrogen burning phase (see Chiosi, Wood \& Capitanio, 1993). In this
scenario the luminosity level is brighter than the canonical case by
0.25 dex.} the PLC relations would provide absolute magnitudes
brighter than the ones obtained with the relations reported in Table 4
by $\sim$ 0.2 mag.

\subsection{Synthetic PL relations}   

PL relations are well known to depend on the topology of the
instability region and on the distribution of pulsators within the
strip. For this reason we did not use the individual models but we
populated the predicted instability strip by adopting the procedure
suggested by Kennicutt et al. (1998) and already used by C00 and
F02. In particular, 1000 pulsators were uniformly distributed from the
blue to the red boundary of the instability strip, with a mass law as
given by $dn/dm=m^{-3}$ over the mass range 5-11$M_{\odot}$ (see C00
for further details).  The resulting synthetic distributions for the
new model sequences are shown in Fig. 5.  Inspection of these plots
confirms the evidence shown in previous papers (Bono et al. 1999b;
C00) that moving toward the shorter wavelengths the $M_{\lambda} -
\log{P}$ distribution of fundamental pulsators with periods longer
than $\sim$ 3 days is much better represented by a quadratic relation
with a clear dependence on both metallicity and the intrinsic width of
the instability strip. On the other hand $J$ and $K$ band
period-magnitude distributions are remarkably narrow, linear and only
slightly dependent on chemical composition.  The resulting synthetic
multiband ($BVRIJK$) PL relations are given in Table 5 and Table 6
(quadratic and linear solutions, respectively) and overplotted as
solid and dashed lines in each panel of the quoted figure.  However,
we wish to remind that, as remarked in C00, the present solutions
refer to a specific pulsator distribution and that different
populations may modify the results. In particular, if the longer
periods ($\log{P}\ge$1.5) are rejected in the final fit, then the
predicted linear $PL$ relations become steeper and the intrinsic
dispersion in the $BVR$ bands is reduced (see Table 7). Such a
selection was also adopted by F02 because the slope of the predicted
linear PL$_V$ and PL$_I$ relations for Cepheids with period
$\log{P}\le$1.5 and the metallicity of the LMC ($Z=0.008$) was found
to be $-2.75\pm0.02$ and $-2.98\pm0.01$, respectively, in very good
agreement with the values ($-2.77\pm$0.03 and $-2.98\pm$0.02) inferred
from the huge sample of LMC Cepheids in the OGLE-II catalog (Udalski
2000).

\section{\bf Theory versus  observations}

In order to test the predictive capabilities of current pulsation
models, in this section we compare the theoretical multiband PL
relations with recent observations for Cepheids belonging to the Milky
Way and the LMC.

In the recent papers by Tammann, Sandage \& Reindl (2003, hereinafter
S03) and Sandage, Tammann \& Reindl (2004, hereinafter S04), accurate
$BVI$ Period-Color (PC), PL and PLC relations are derived on the basis
of large databases for Galactic and LMC Cepheids, respectively. These
authors show that the PL relations for Cepheids in the Galaxy, LMC and
SMC have significantly different slopes (S03), and in particular S04
report about the experimental evidence of a change of the slope (a
break) of the PL relation for the LMC Cepheids near 10 days. This last
result is supported by the broken PC relation for the LMC Cepheids
showed in S04. Indeed, any nonlinearity of the PC relations must be
reflected in the PL relation.  These results represent an important
tool to test the accuracy of current model predictions concerning both
the nonlinearity of optical PL relations and the dependence of Cepheid
properties on metal abundance.

As for the first point, the theoretical evidence for the nonlinearity
of $BVI$ PL and PC relations was already reported in our previous
papers (see e.g. Bono et al. 1999b, C00) and for this reason we
usually adopt linear PL relations for $\log{P}\le 1.5$ (see previous
section). In particular, Fig. 4 in C00 shows that the nonlinear
behavior of the PC relation is more evident for $Z=0.004$ and
$Z=0.008$ (representative of the LMC and SMC metallicity
respectively), whereas it is significantly reduced for $Z=0.02$
(representative of the metallicity of Galactic Cepheids). This result
is in agreement with S03 and S04 that found the break at 10 days only
for the LMC Cepheids. On this basis, in order to compare our results
with the data presented by S04, we also derived the theoretical linear
PL relations for $\log{P}\le 1.0$ and $\log{P} > 1.0$, for the metal
abundances $Z=0.004$ and $Z=0.008$. These theoretical relations are
compared with the S04 LMC sample in Fig. 6.  In this plot dots
represent the OGLE sample selected as in S04 and open circles the
longer period Cepheids collected by S04 from several sources (see S04
for details). Solid lines show the fits for $\log{P}<1.0$ and
$\log{P}>1.0$ obtained by S04 (by using both samples), whereas short
and long dashed lines represent the theoretical PL relations for
$Z=0.004$ and $Z=0.008$ respectively, with the same period selection.
We also plotted the theoretical PL relations for $Z=0.004$ in order to
take into account the significant metallicity dispersion of LMC
Cepheids (Luck et al. 1998) and the results by Bono et al. (1999b)
that at longer periods the observed distribution of LMC Cepheids in
the period-magnitude diagram is better represented by the theoretical
one for $Z=0.004$ (see Bono et al. 1999b for details). Inspection of
Fig. 6 and of Tables 7 and 8 suggests that the slopes obtained for
$Z=0.004$ and $Z=0.008$ in the two period ranges are similar. However,
the $B$ and $V$ PL relations for $Z=0.008$ are systematically fainter
than the observational one, which shows a better agreement with the
model predictions for $Z=0.004$. On the other hand, in the $I$ band we
have a very good agreement between the empirical fits and the
relations for $Z=0.008$, whereas the relations for $Z=0.004$ seem to
be systematically brighter.

In order to investigate the dependence of Cepheid properties on
metallicity, in Fig. 7 we plot the S04 Galactic sample with our PL
relations for $Z=0.02$ and $Z=0.01$. Open circles represent the S04
sample with distances from Baade-Becker-Wesselink expansion parallaxes
and filled circles indicate the S04 sample with distances from
cluster/associations Cepheids (see S04 for details). The solid line
represents the S04 fit (obtained by using both the samples, see S04
for details), the long dashed line is our PL relation (for $\log{P}\le
1.5$) for $Z=0.01$, $Y=0.26$ and the other lines are the PL relations
(for $\log{P}\le 1.5$) for $Z=0.02$ and the labeled values of the
Helium content. Our theoretical relations are flatter than the S04
one, but a better agreement with the data is found when we consider the
model predictions for $Z=0.01$, $Y=0.26$ and $Z=0.02$ and
$Y=0.25$,0.26, at least for $\log{P} \le 1.5$. For the other chemical
compositions, the discrepancy is particularly evident at the longer
periods and for the $B$ and $V$ bands. In this context, we remind that
current comparisons rely on the assumption of static model atmospheres
and that theoretical colors can be affected by systematic
uncertainties. Moreover, the $B$ and $V$ PL relations are very
sensitive to the topology of the instability strip and, in turn, to the
adopted input physics and to the treatment of convection in the
pulsation models (see below). The effects of these uncertainties are
likely more important for long period expanded structures.  The better
agreement obtained for models with $Z=0.01$ supports the recent
suggestions by Asplund, Grevesse \& Sauval (2005) that the solar
metallicity is lower ($Z \simeq$ 0.01) than usually adopted
($Z=0.02$).

Finally, we consider the interferometric results for seven Galactic
Cepheids by Kervella et al. (2004a,b,c), which are a subsample of the
S04 dataset, but only with $V$ and $K$ band observations. These
authors have presented accurate radius and distance determinations
based on interferometric measurements of the angular diameter.  On
this basis they have derived new period-radius, as well as $V$ and
K band PL relations (Kervella et al. 2004b, hereinafter K04b) by
assuming the slopes by Gieren et al. (1998). All these objects have
metal abundance close to the solar one. For $l$Car the metal abundance
reported by K04b is about twice the solar value, but the recent
spectroscopic measurement by R05 suggests a solar value also for this
object. In Fig. 8 we show the comparison between our predicted $K$ band
(upper panel) and $V$ band (lower panel) PL relations (for $\log{P} \le$1.5,
see Table 7) at solar metallicity with the labeled helium abundances
and the interferometric results by K04b. The intrinsic dispersion of
the theoretical relations is represented by the vertical error bar in
the labels, whereas the solid line represents the empirical PL
relation obtained by fitting the data. For $\eta$Aql ($\log{P} = 0.8559$)
we have reported both the determinations used by K04b. Moreover, for
variable $l$Car, the revised magnitude values by K04c, based on a more
accurate interferometric determination of radius and distance, are
also reported (empty circle)\footnote{In C05 we discussed the
possibility that $l$Car is a peculiar variable star. Indeed, on the
basis of the comparison between the pulsational and evolutionary
masses, it seems to be an object on the first crossing of the
instability strip.}.  An inspection of this figure shows that in the $K$
band our theoretical relations are able to reproduce the data within
the errors, whereas in the $V$ band the predicted PL relations are
fainter than the empirical one and fail to match the location of
the two pulsators with the smallest error on the absolute
magnitude. In Fig. 9, we also compare K04b data with the theoretical
relation at $Z=0.01$. We notice that, as already found for the
comparison with S04 data, model predictions at this lower metal
abundance better reproduce the interferometric results for Galactic
Cepheids.

The discrepancies found in the comparison with the
observational data by S04 and K04abc can be due, at least in part, to
the uncertainties still affecting the adopted theoretical scenario. In
particular we remind that current models are based on specific
assumptions concerning both the evolutionary ML relations (see Bono et
al. 1999a, 2000c, C05 for details) and the value of the mixing length
($\alpha$) parameter adopted in the treatment of convection to close
the system of nonlinear dynamical and convective equations (see Bono
\& Stellingwerf 1994, Bono et al. 1999a).  For the former point we
refer the interested reader to the detailed discussion by C05. As for
the mixing length parameter, even if recent results based on the
theoretical fitting of observed Cepheid light curves suggest that the
value of the $\alpha$ parameter should increase when moving from the
blue to the red boundary of the instability strip (see Bono,
Castellani \& Marconi 2002a), in agreement with recent results
obtained from the modeling of RR Lyrae stars (see e.g. Di Criscienzo,
Marconi \& Caputo 2004), all the models presented and adopted in this
paper have been computed with $\alpha=1.5$. Specific model sets at
$Z=0.02$ $Y=0.28$ and $Z=0.01$ $Y=0.26$ computed, by increasing $\alpha$ from
1.5 to 1.8, show that the instability strip gets significantly
narrower with the red boundary getting bluer by at least 300-400 $K$ and
a smaller redward shift of the blue boundary. This occurrence is due
to the higher sensitivity of the red part of the instability strip to
the efficiency of the convective transfer. On the basis of these
results and taking into account the possibility that the $\alpha$
value is different at the blue and red edge of the strip, we expect
that the PL relation may become brighter and steeper when $\alpha$
increases. 

\section{Metallicity and helium effects on the predicted distance
scale} In order to test the results presented by F02 concerning the
combined metallicity and helium effects on the Cepheid distance scale
and provide a refined theoretical correction, we applied the same
procedure adopted by the quoted authors to our extended model set. In
particular, we considered our models with the various chemical
composition as real Cepheids at the fixed distance modulus $\mu_0
=0$~mag. By applying the predicted linear $V$ and $I$ band PL
relations with $Z=0.008$ and $Y=0.25$ for $\log{P}\le{1.5}$, we determined
the value $\mu_{0,0.008}$ for all the pulsators. This method simulates
the HST Key project procedure (e.g. F01) which uses observations in
the two bands $V$ and $I$ and adopts the LMC PL relations as universal
(F01, Udalski et al. 1999). In this context, we adopted
$\mu_V-\mu_I=E(V-I)$ and $A_I/E(V-I)=1.54$ from Cardelli et
al. (1989). The derived $\mu_{0,0.008}$ values confirm the results by
F02: a) for periods shorter than 10 days, the discrepancy between
$\mu_{0,0.008}$ and the real value ($\mu_0=0$~mag) is small enough ($<
0.1$ mag) to support the adoption of universal LMC-referenced PL
linear relations; b) for periods longer than 10 days, the discrepancy
is larger than $0.1$ mag (up to 0.3 mag for $Z=0.02$ and $Y=0.28$ and
period longer than 20 days) over the range $Z \sim 0.01 - 0.04$, so
that a correction is required. In particular a dependence on chemical
composition of the form suggested by F02 for longer period Cepheids is
also found on the basis of the extended model set presented in this
paper.  The mean correction for $\log P \ge 1.0$ is $$c=-3.642 +
11.511 Y - 1.697 \log{Z} + 5.334 Y \log{Z}$$ whereas for Cepheid samples
with $\log P \ge 1.3$ the mean correction is better reproduced by:
$$c=-5.894 + 18.141 Y - 2.792 \log{Z} + 8.576 Y \log{Z}$$ both with an
intrinsic uncertainty of $\pm 0.02$mag and by assuming
$\Delta{Y}/\Delta{Z} > 1.5$ (see Fig. 10).  For $\Delta{Y}/\Delta{Z}
\le 1.5$ the dependence is more complicated but the correction is
always lower than 0.1 mag and can be neglected as in the case of
shorter periods. We also remind that on the basis of current estimates
in the literature, we do not expect such low values for the
$\Delta{Y}/\Delta{Z}$ parameter (Izotov \& Thuan, 2004 and references
therein).  As shown in Fig. 10, the predicted mean metallicity
correction implies that pulsators get fainter as their metallicity
increases until a turnover point is reached close to the solar metal
abundance. Such a behavior was already mentioned to be in agreement
with the recent spectroscopic results by R05 and to reproduce the
empirical metallicity correction by Kennicutt et al. (1998) for
$\Delta{Y}/\Delta{Z} \sim 3.5$. We also notice that the mean
correction gets smaller when the lowest period of the investigated
sample decreases from 20 (lower panel) to 10 days (upper panel). In
particular, the mean correction for $\log{P} \ge 1$ is higher than 0.1
mag only at the highest metallicities ($Z \ge 0.03$) and
$\Delta{Y}/\Delta{Z}$ values ($\ge 3$). On the other hand, if the
Cepheid periods are longer than or equal to 20 days, the mean
correction is larger than 0.1 mag on a wide range of metallicities and
helium contents.

\section{Conclusions}    

We have presented an extended set of nonlinear convective pulsation
models at varying the metallicity and $\Delta{Y}/\Delta{Z}$ ratio. On
this basis, we obtain the following main results:

\begin{enumerate} 

\item 
we have confirmed our previous results concerning the shift of the
instability strip toward lower effective temperatures as the metal
abundance increases at fixed $\Delta{Y}/\Delta{Z}$, at least up to
$Z=0.03$. At the same time, when passing from $Z=0.03$ to $Z=0.04$,
the strip narrows due to the reduced efficiency of pulsation. The
effect of variation of the helium abundance at fixed metallicity is
lower than the one obtained by varying the metallicity at fixed
$\Delta{Y}/\Delta{Z}$. In particular, the fundamental red edge
slightly moves toward higher effective temperatures as the helium
content increases, whereas the fundamental blue edge does not show a
clear trend;

\item
inspection of the bolometric light curves, in the period range
affected by the HP phenomenon, shows that passing from $Z=0.01$ to
$Z=0.04$ the period corresponding to the HP center moves from
$\sim$10.5~d to $\sim$8.2~d, in agreement with the empirical evidence
of a decrease of PHP as the metallicity increases and with our
previous theoretical results for $Z\le 0.02$; 

\item 
a comparison with the large database of Galactic and LMC Cepheids by
Sandage et al. and Tamman et al. shows that, in agreement with the
conclusions of these authors, the $BVI$ PL relations for LMC pulsators
are well reproduced by linear theoretical relations with a break at
$\log{P} = 1$. As for the dependence on metallicity, we find that our
theoretical PL relations for $Z=0.02$ are generally flatter than the
empirical ones for Galactic Cepheids, with the discrepancy increasing
toward the longer periods. A good agreement is obtained when, on the
basis of suggestions in the recent literature, $Z=0.01$ is assumed as
solar metal abundance in the models;

\item 
a comparison with recent accurate interferometric results by Kervella
et al. shows that in the $K$ band our theoretical period-luminosity
relations are able to reproduce the data within the errors, whereas in
the $V$ band the predicted PL relations are fainter than the empirical
one. Among the possible reasons for such a discrepancy, as well as for
the one quoted in the previous point, we have identified the
uncertainty on the mixing length parameter adopted in the treatment of
convection and on the value of the solar metallicity;

\item 
we have derived the theoretical correction to the distance moduli
inferred with the HST Key project procedure, when the effect of
metallicity and helium abundance are taken into account. We find that
this effect is smaller than 10\% and can be neglected for Cepheid
samples with $\log{P} < 1.0$ and for $\Delta{Y}/\Delta{Z} \le 1.5$. For
longer periods and higher $\Delta{Y}/\Delta{Z}$ values, the dependence
of the mean theoretical correction on chemical composition has the
same analytical form of the one found by F02, but it is shown to become
important only for periods longer than 20 days. 

\end{enumerate}

\acknowledgments  
It is a pleasure to thank Filippina Caputo for a critical reading of
the manuscript, useful discussions and suggestions. We also thank
G. Tammann and B. Reindl for sending us the Galactic and LMC Cepheid
samples adopted in the paper by Sandage, Tammann \& Reindl (2004) and
an anonymous referee for his useful comments.  Financial support for
this study was provided by MIUR, under the scientific projects
``Stellar Populations in the Local Group'' (P.I.: Monica Tosi), 
``Continuity and Discontinuity in the Milky Way Formation'' (P.I.:
Raffaele Gratton) and ``On the evolution of stellar systems:
fundamental step toward the scientific exploitation of VST''
(P.I. Massimo Capaccioli). This project made use of computational
resources granted by the ``Consorzio di Ricerca del Gran Sasso''
according to the ``Progetto 6: Calcolo Evoluto e sue applicazioni
(RSV6) - Cluster C11/B''.

\pagebreak

\pagebreak    
    

\begin{table}
\caption{Empirical and theoretical PL relations for LMC
Cepheids with the break at $\log{P} = 1$}
\begin{center}
\begin{tabular}{c|c|cc|cc}
\hline
\hline
&&&&&\\[-10pt]
{\bf Sample}    & {\bf Band} & \multicolumn{2}{c}{\bf slope} &
\multicolumn{2}{c|}{\bf zero
point} \\
          &      & $\log{P} < 1.0$ & $\log{P} > 1.0$ & $\log{P} < 1.0$ & $\log{P}> 1.0 $\\[5pt] 
\hline 
&&&&&\\[-12pt]
 S04      & $B$ & $-2.68 \pm 0.08$ & $-2.15 \pm 0.13$ & $-0.99 \pm 0.05$ &$-1.40 \pm 0.18$ \\ [2pt]
$Z=0.004$ & $B$ & $-3.09 \pm 0.04$ & $-2.04 \pm 0.11$ & $-0.64 \pm 0.03$ &$-1.68 \pm 0.13$ \\ [2pt]
$Z=0.008$ & $B$ & $-2.75 \pm 0.05$ & $-1.84 \pm 0.11$ & $-0.72 \pm 0.03$ &$-1.64 \pm 0.14$ \\ [2pt]
\hline
&&&&&\\[-12pt]
 S04      & $V$ & $-2.96 \pm 0.06$ & $-2.57 \pm 0.10$ & $-1.33 \pm 0.04$ &$-1.63 \pm 0.13$ \\ [2pt]
$Z=0.004$ & $V$ & $-3.21 \pm 0.03$ & $-2.47 \pm 0.08$ & $-1.13 \pm 0.02$ &$-1.87 \pm 0.10$ \\ [2pt]
$Z=0.008$ & $V$ & $-2.97 \pm 0.03$ & $-2.32 \pm 0.08$ & $-1.22 \pm 0.02$ &$-1.87 \pm 0.10$ \\ [2pt]
\hline
&&&&&\\[-12pt]
 S04      & $I$  & $-3.10 \pm 0.04$ & $-2.82 \pm 0.08$ & $-1.85 \pm 0.02$ &$-2.08 \pm 0.11$ \\ [2pt]
$Z=0.004$ & $I$  & $-3.30 \pm 0.02$ & $-2.79 \pm 0.05$ & $-1.79 \pm 0.02$ &$-2.30 \pm 0.07$ \\ [2pt]
$Z=0.008$ & $I$  & $-3.13 \pm 0.02$ & $-2.68 \pm 0.05$ & $-1.84 \pm
0.02$ &$-2.30 \pm 0.07$ \\ [2pt] 
\hline
\end{tabular}
\end{center}
\end{table}

\clearpage   

\def\fnum@figure{{\rm Fig.\space\thefigure.---}}
\figcaption{The theoretical instability strip as a function of metallicity for two different assumptions on the $\Delta{Y}/\Delta{Z}$ ratio.}   


\def\fnum@figure{{\rm Fig.\space\thefigure.---}}
\figcaption{The theoretical instability strip as a function of the helium abundance for three different assumptions on the metal content.}   


\makeatletter
\def\fnum@figure{\rm Fig.\space\thefigure a.---}
\makeatother

\figcaption{Bolometric light curves of models with $Z=0.01$, $Y=0.26$. The mass and luminosity values are labeled in the first column plots, whereas the model effective temperature and period is reported in each panel;}


\setcounter{figure}{2}
\makeatletter
\def\fnum@figure{\rm Fig.\space\thefigure b.---}
\makeatother

\figcaption{The same as Fig. 3a but for Z=0.02 Y=0.25}   


\setcounter{figure}{2}
\makeatletter
\def\fnum@figure{\rm Fig.\space\thefigure c.---}
\makeatother

\figcaption{The same as Fig. 3a but for Z=0.02 Y=0.26}   


\setcounter{figure}{2}
\makeatletter
\def\fnum@figure{\rm Fig.\space\thefigure d.---}
\makeatother

\figcaption{The same as Fig. 3a but for Z=0.03 Y=0.275}   


\setcounter{figure}{2}
\makeatletter
\def\fnum@figure{\rm Fig.\space\thefigure e.---}
\makeatother

\figcaption{The same as Fig. 3a but for Z=0.03 Y=0.335}   


\setcounter{figure}{2}
\makeatletter
\def\fnum@figure{\rm Fig.\space\thefigure f.---}
\makeatother

\figcaption{The same as Fig. 3a but for Z=0.04 Y=0.25}    


\setcounter{figure}{2}
\makeatletter
\def\fnum@figure{\rm Fig.\space\thefigure g.---}
\makeatother

\figcaption{The same as Fig. 3a but for Z=0.04 Y=0.29}

\makeatletter
\def\fnum@figure{\rm Fig.\space\thefigure.--- }
\makeatother

\figcaption{Enlarged portion of the light curve atlas reported in the previous
figures for model sets showing evidence of the HP phenomenon. The
chemical composition is reported in the first column plots, together
with the luminosity level.}  


\figcaption{
Synthetic multiband PL relations at varying the chemical composition
(see text for details). The solid and dashed lines represent the
linear and quadratic regression respectively.}


\figcaption{Comparison between theoretical PL relations with the break
at $\log{P} = 1$ and LMC Cepheids by S04.  Dots represent the OGLE
sample and open circles a sample of longer period Cepheids (see text
for details). Solid lines show the fits for $log{P}<1.0$ and
$log{P}>1.0$ obtained by S04, whereas short and long dashed lines
represent the theoretical PL relations for $Z=0.004$ and $Z=0.008$
respectively.}


\figcaption{Comparison between theoretical PL relations and the S04
Galactic sample.  Open and filled circles represent the S04 sample
with distances from Baade-Becker-Wesselink expansion parallaxes and
cluster/associations Cepheids respectively. The solid line represents
the S04 fit (see text for details), the long dashed line is our PL
relation (for $log{P}\le 1.5$) for $Z=0.01$ and $Y=0.26$ and the other
lines are the PL relations (for $log{P} \le 1.5$) for $Z=0.02$ and
different values of the Helium content (see labels).}


\figcaption{Comparison between the predicted $V$ (lower panel) and $K$ (upper panel)  PL relations at $Z=0.02$, and the labeled helium abundances, and the interferometric results by Kervella et al. (see text for details). The solid line is the empirical linear regression to the data.}   


\figcaption{The same as in Fig. 6 but with the theoretical PL relations for $Z=0.02$, $Y=0.28$ and $Z=0.01$, $Y=0.26$.}   


\figcaption{Upper panel: predicted mean metallicity correction for the labeled $\Delta{Y}/\Delta{Z}$ ratios and periods longer than 10 days; Lower panel: the same but for periods longer than 20 days.}

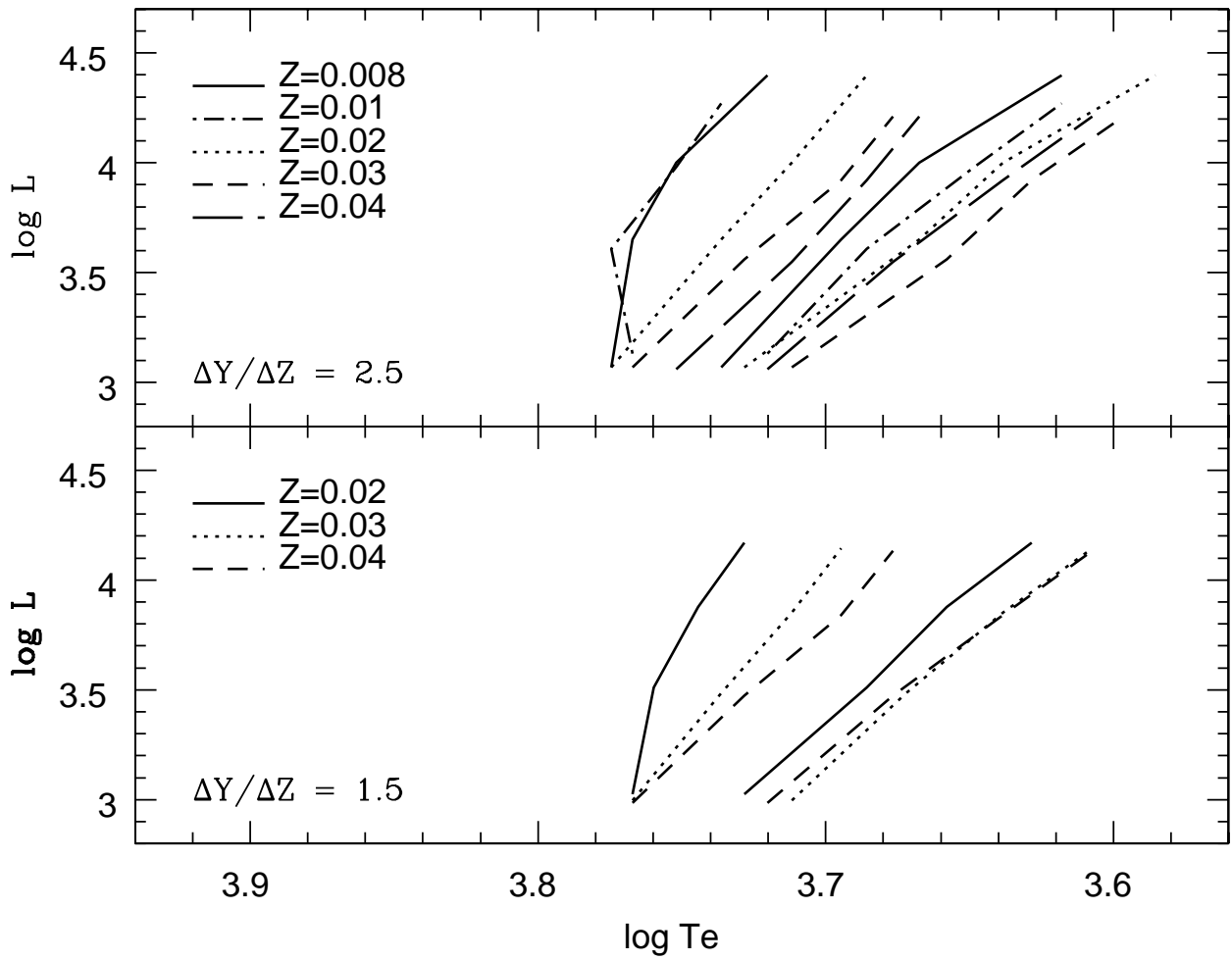
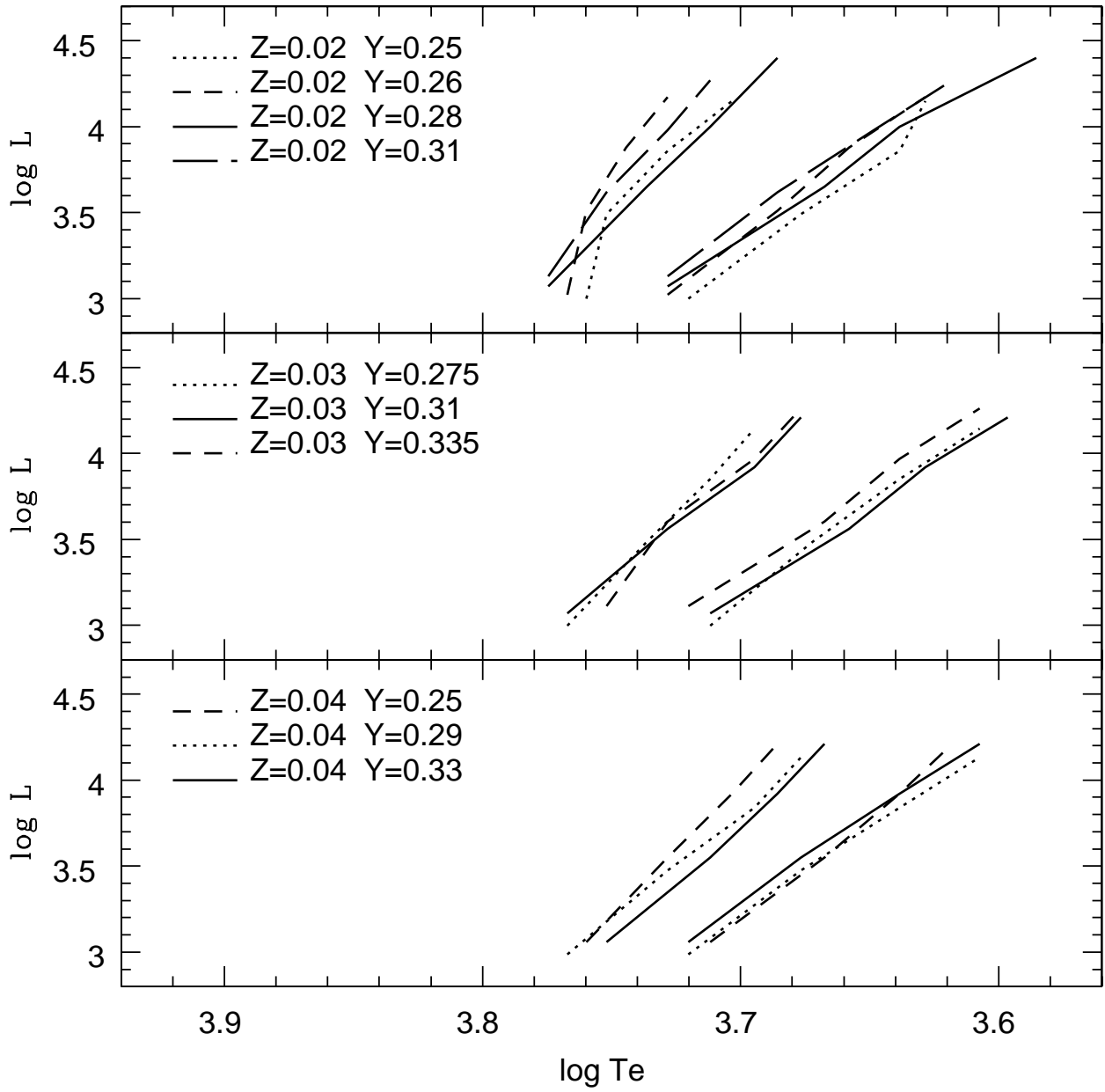
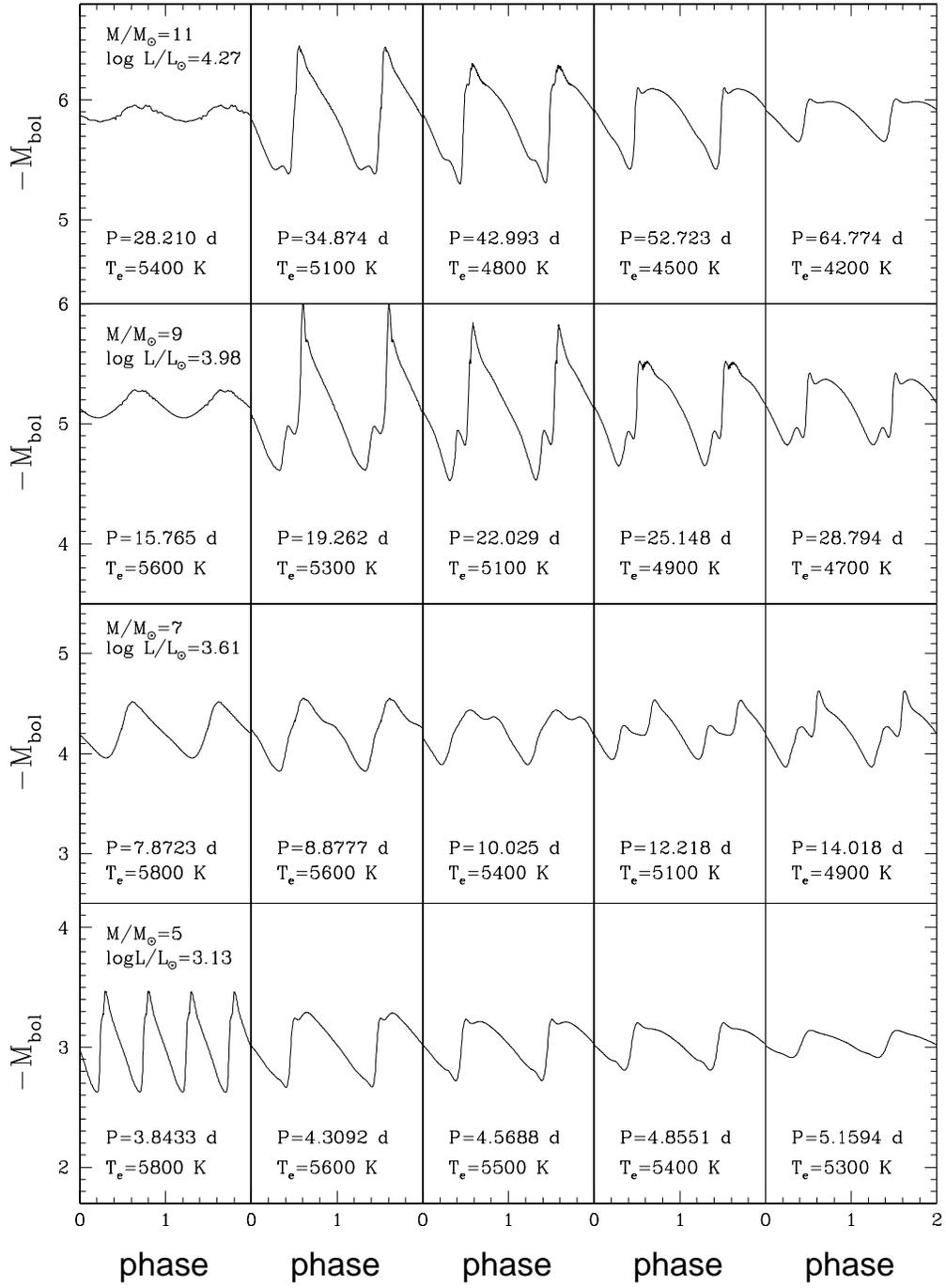
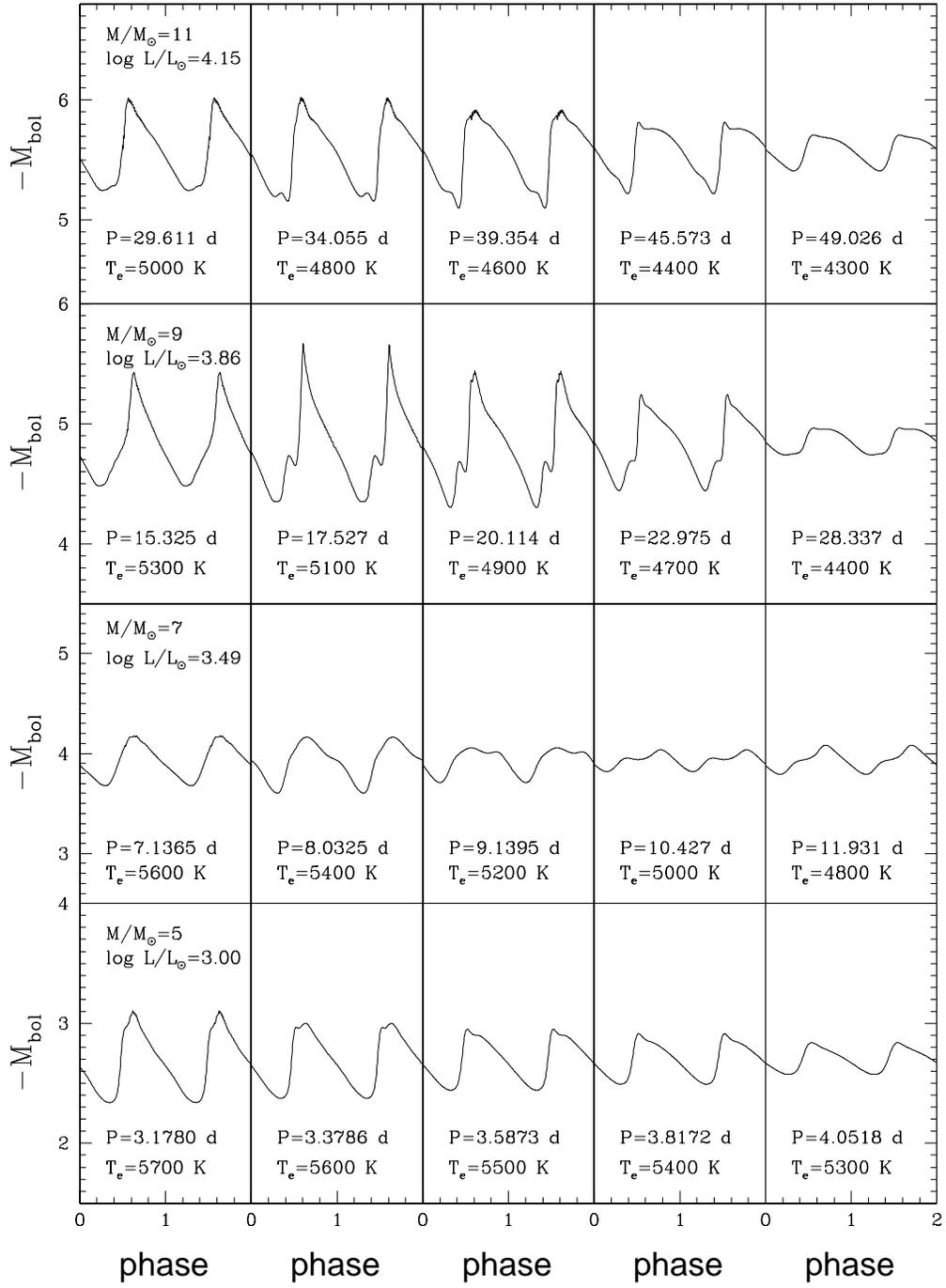
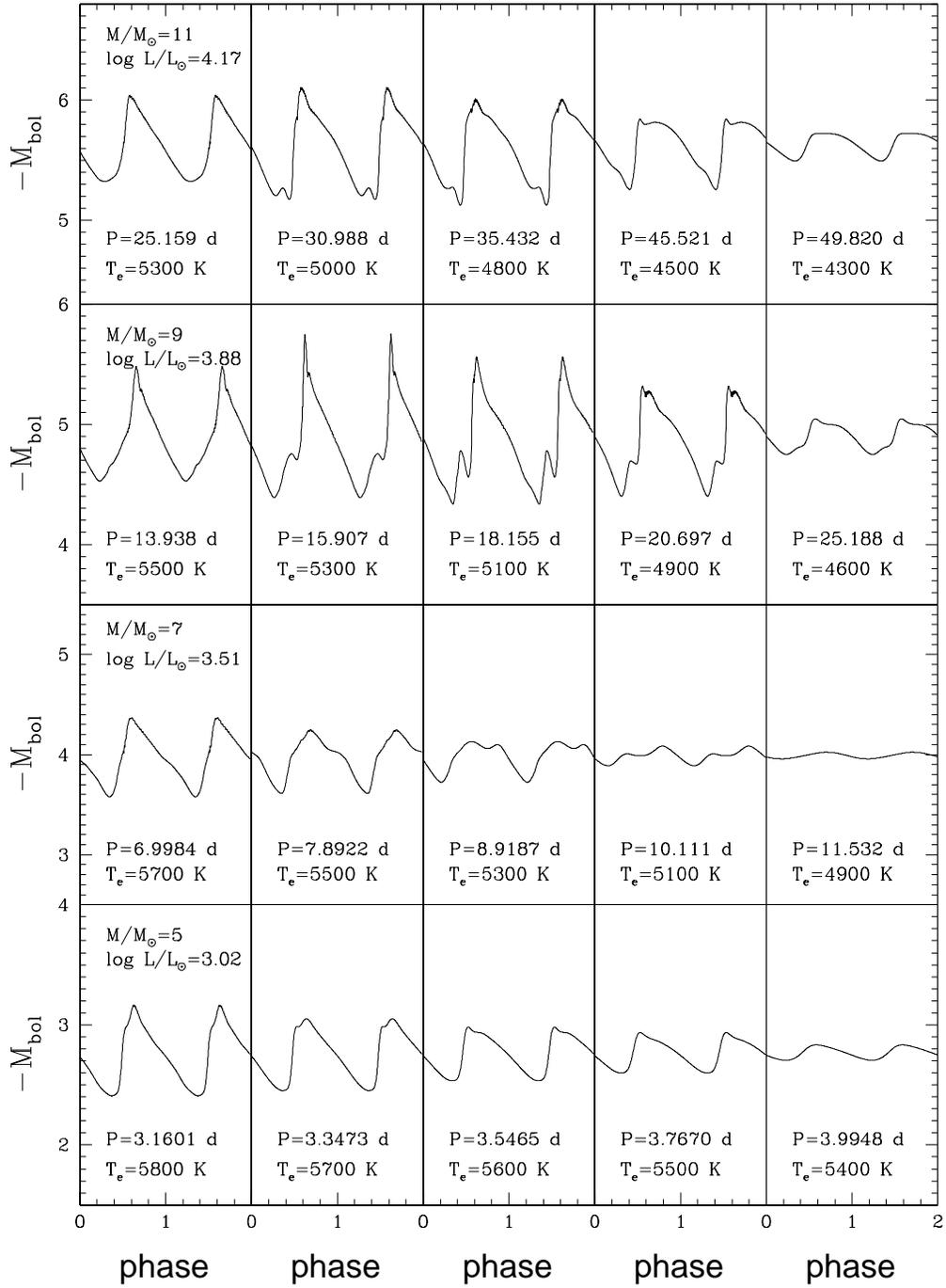
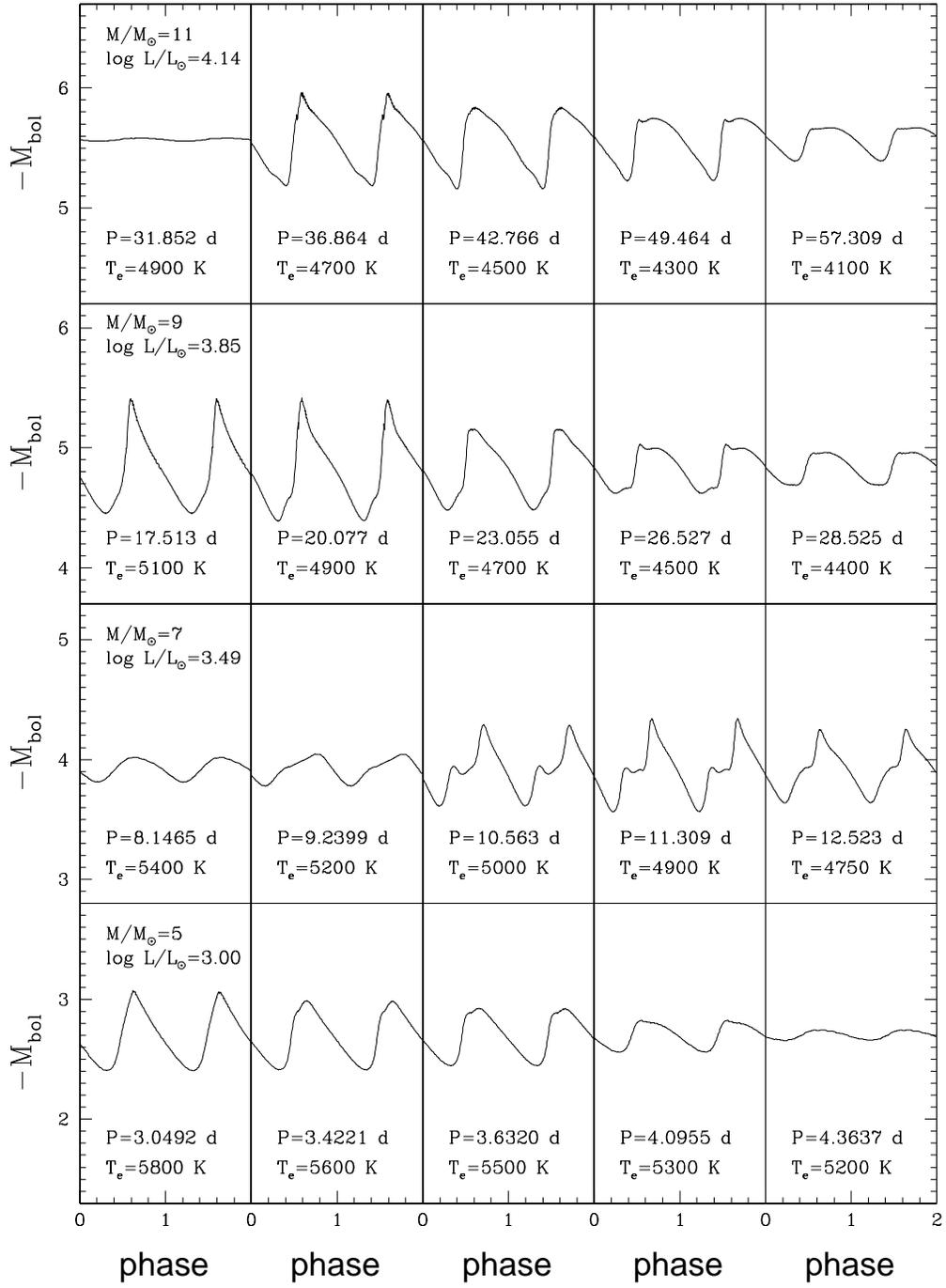
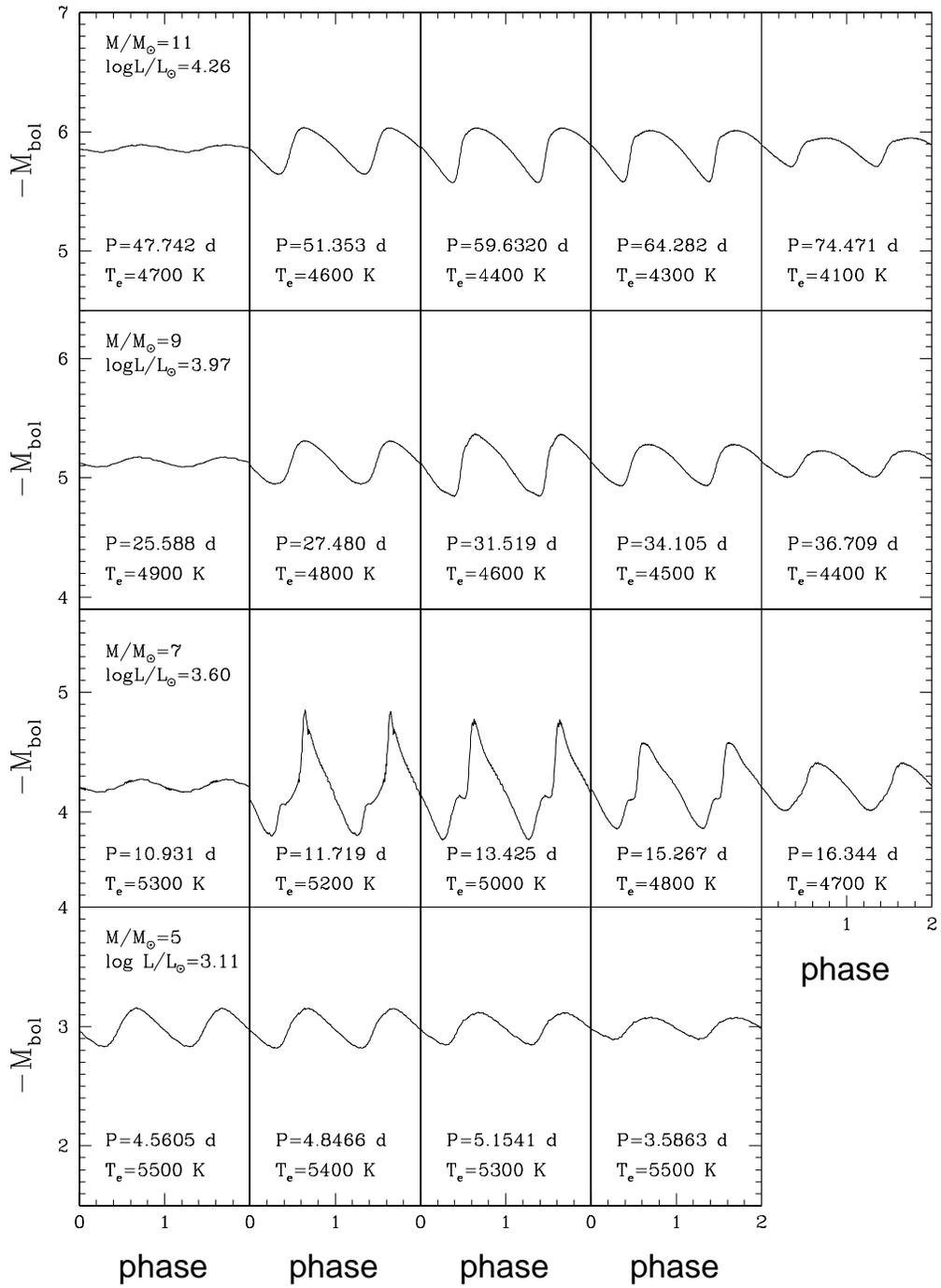
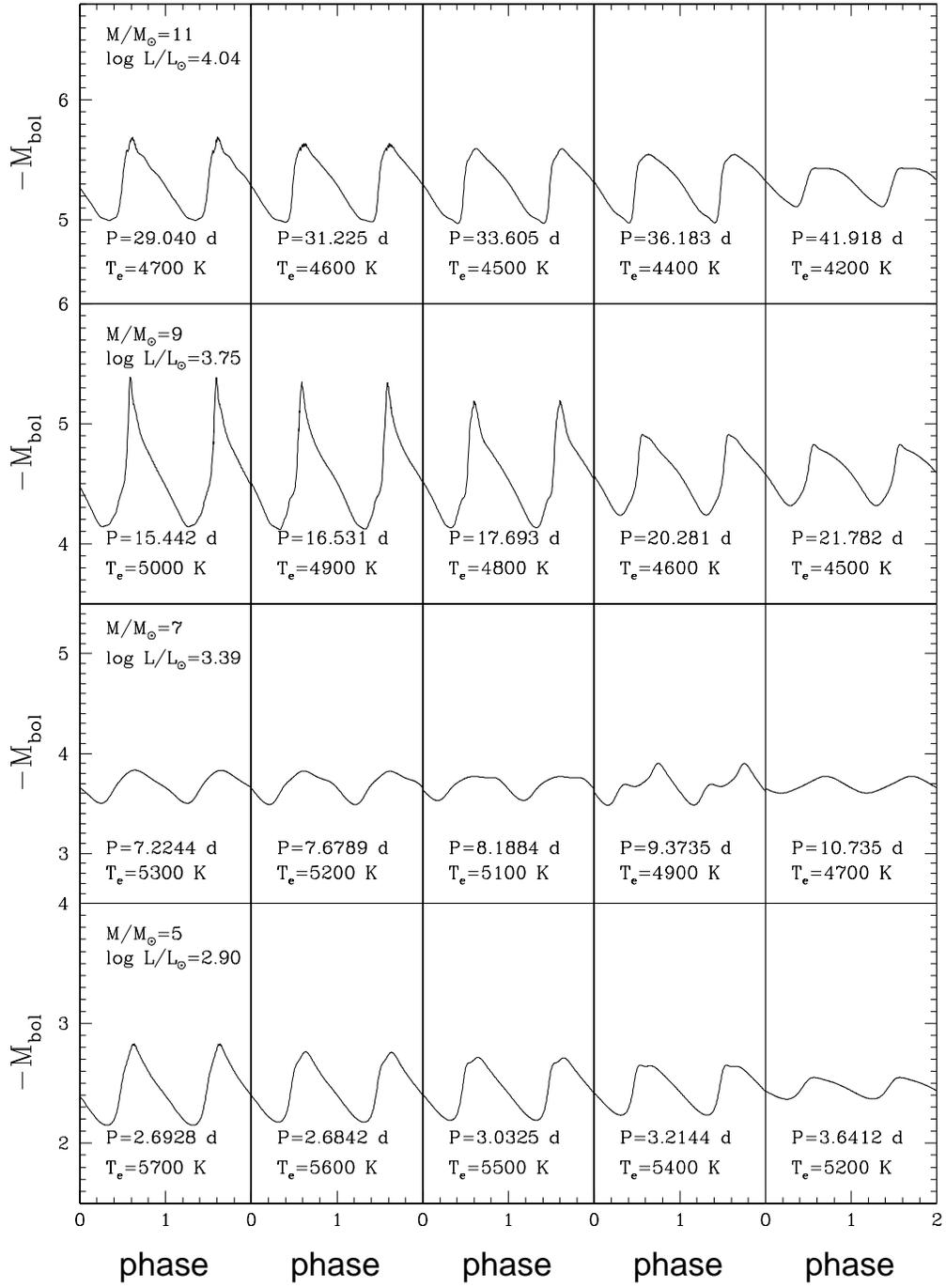
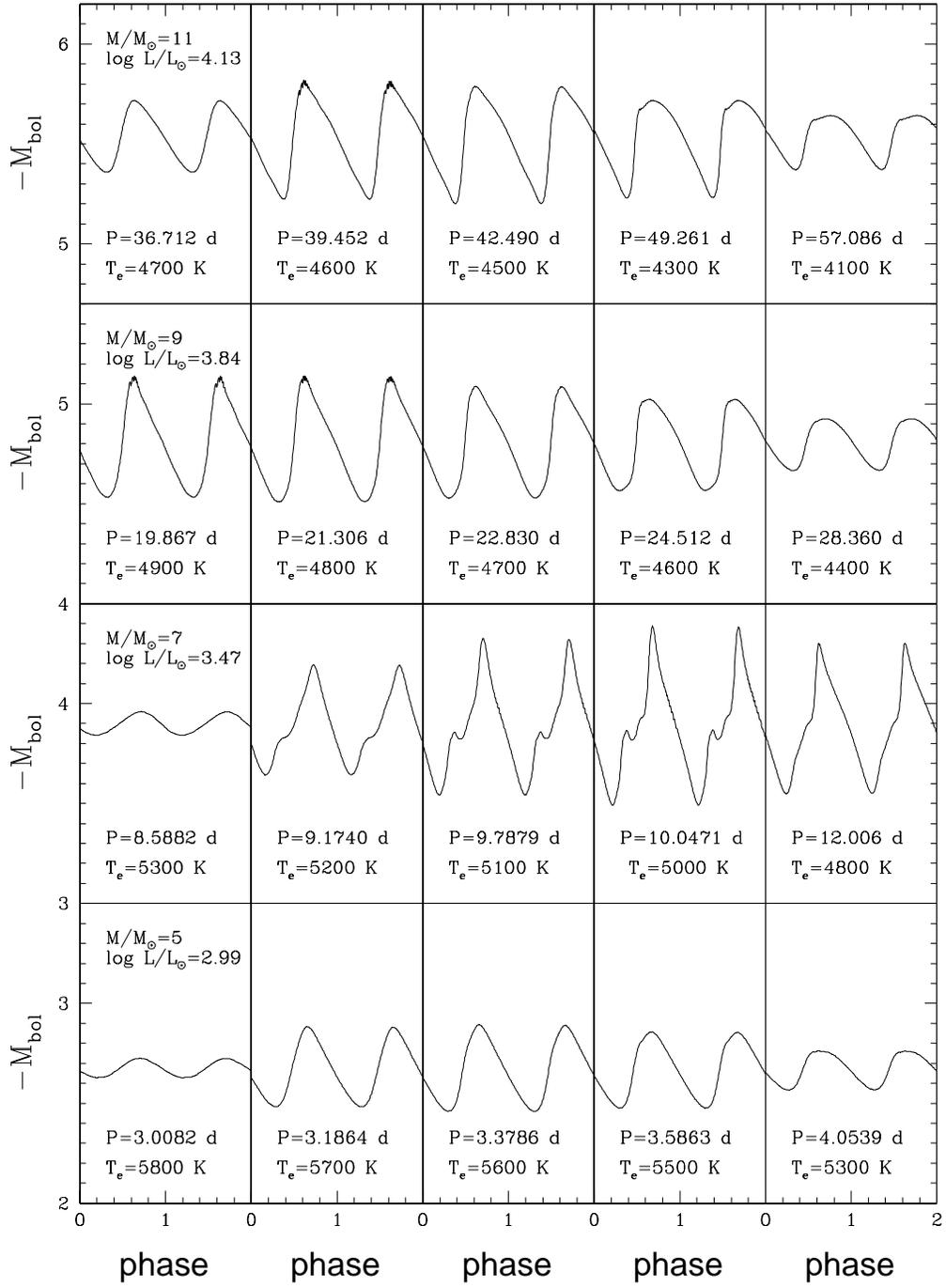
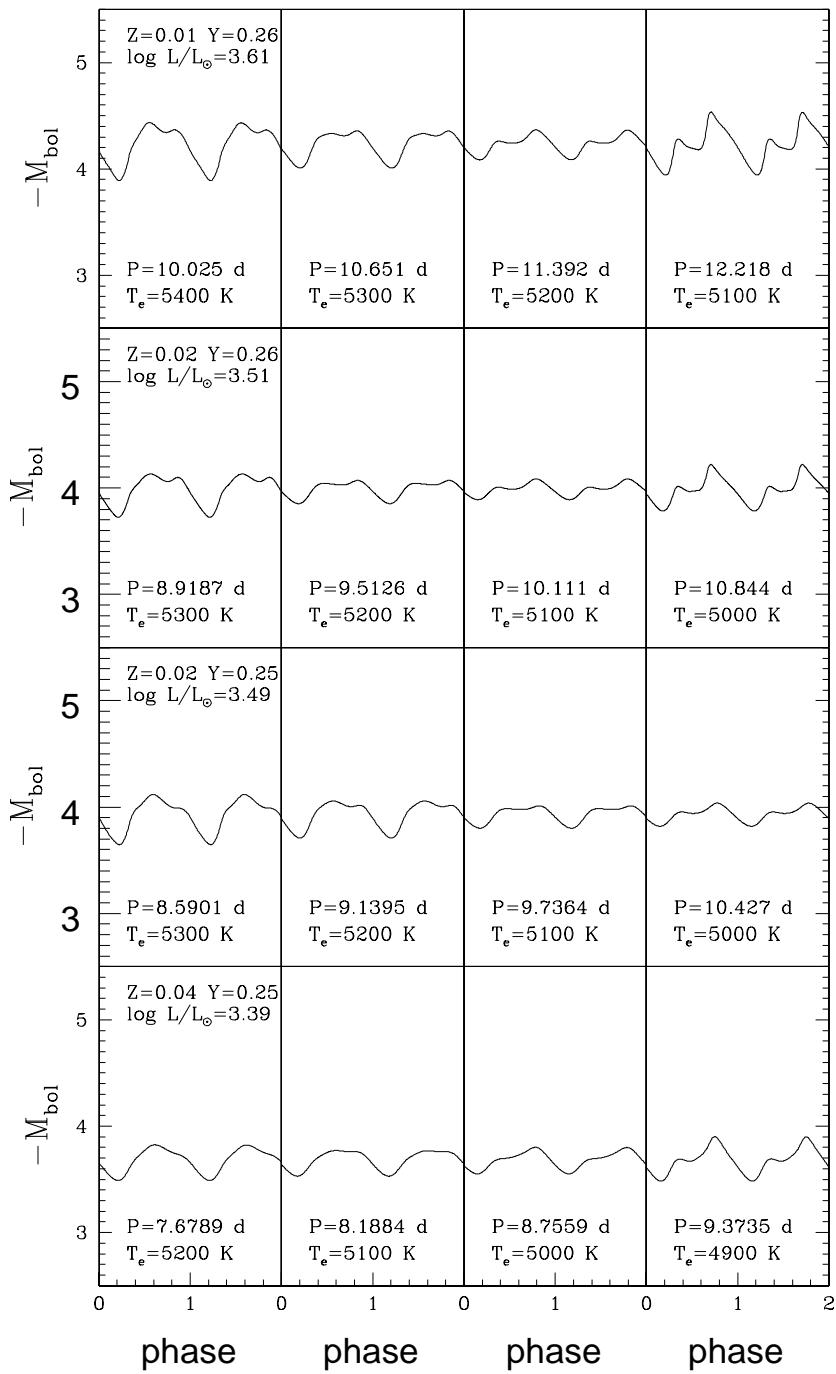
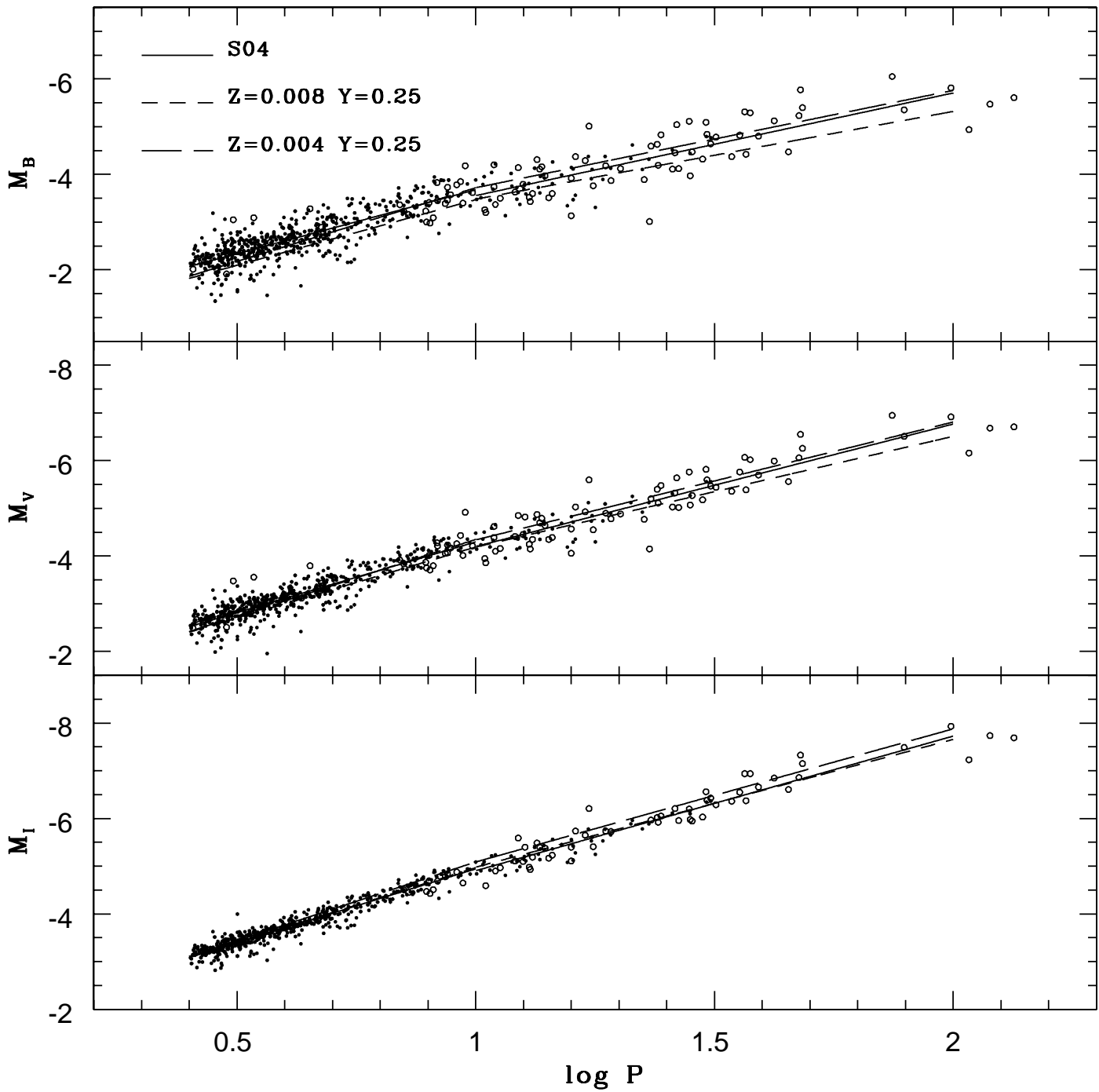
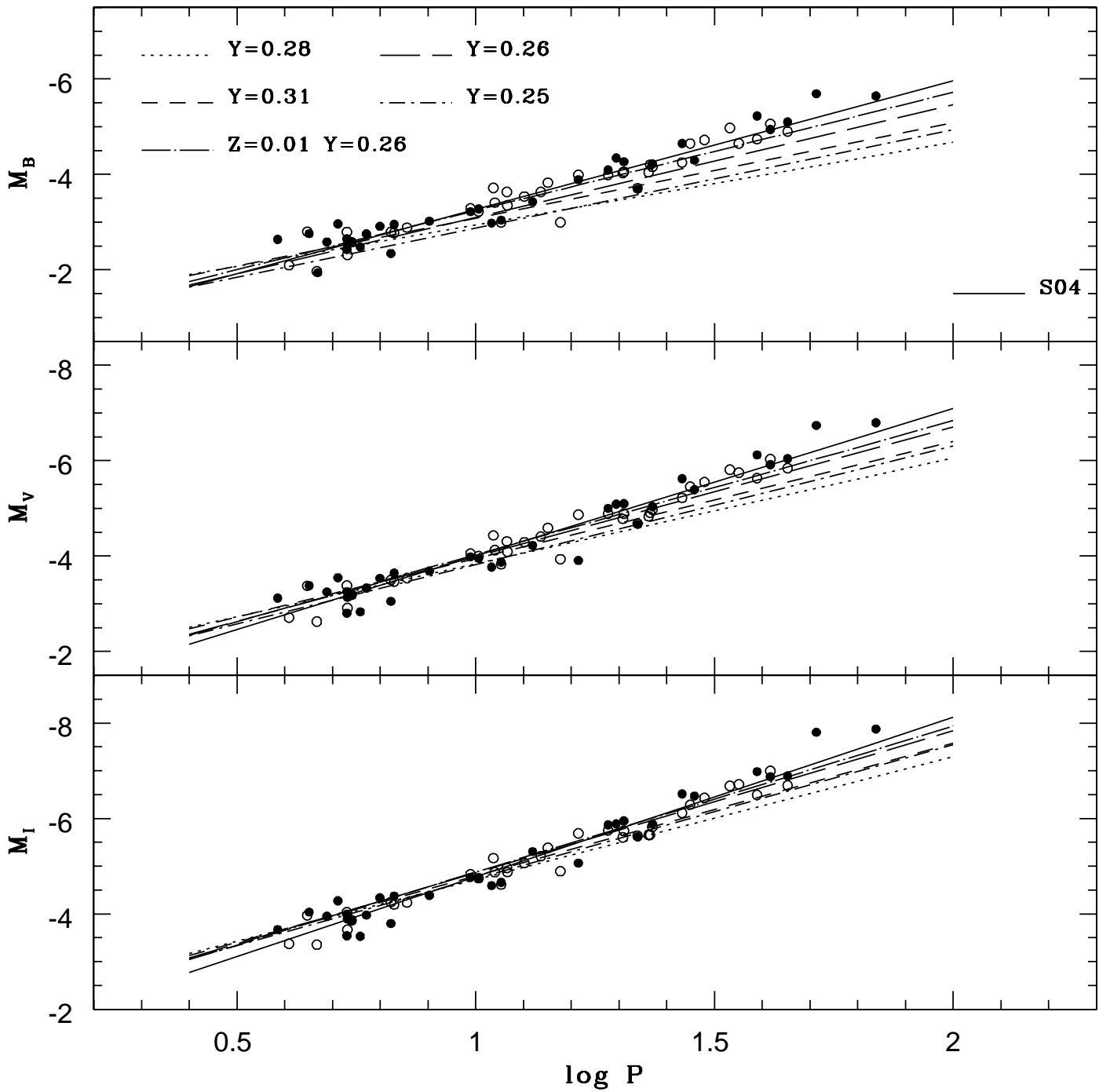
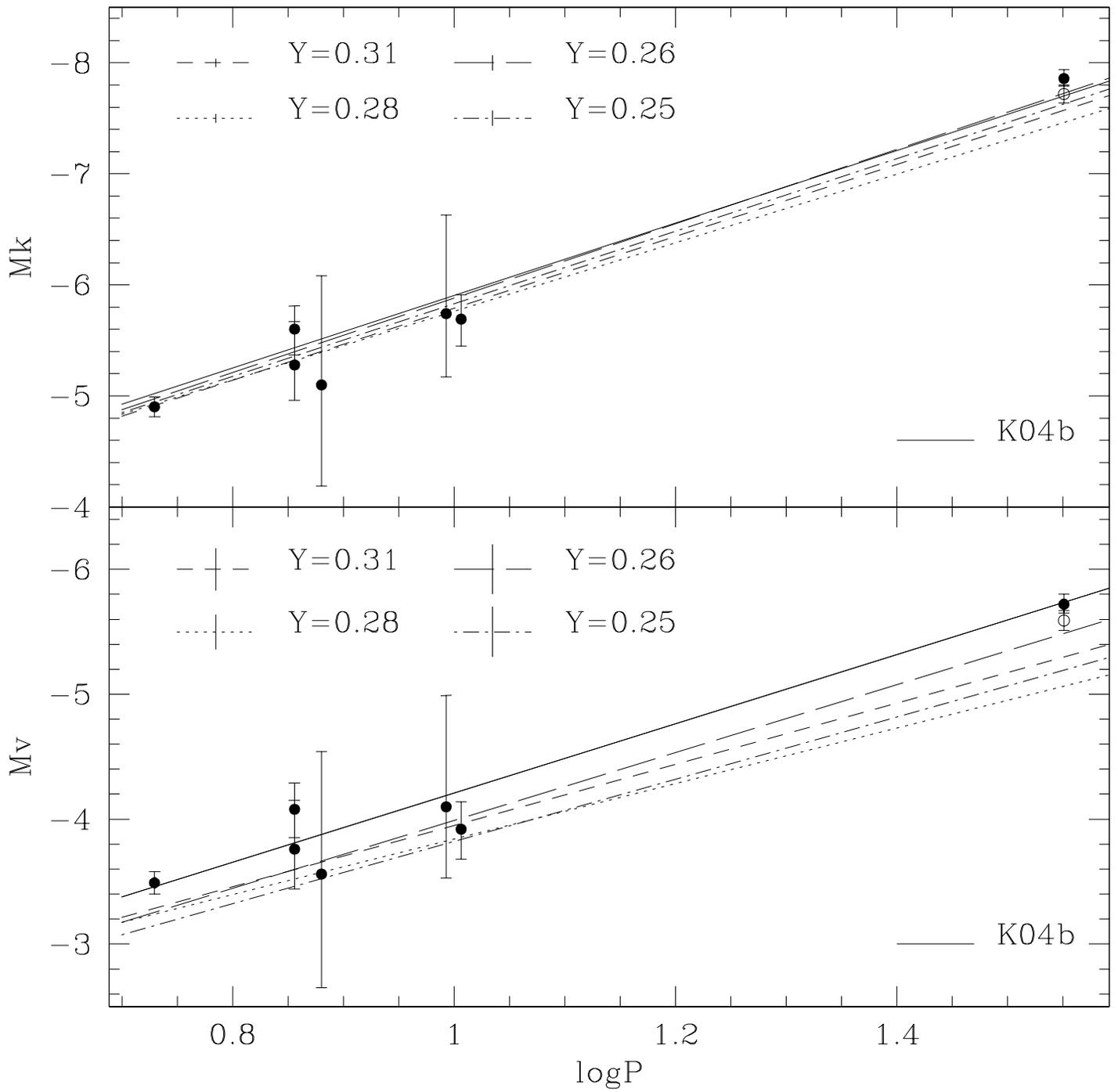
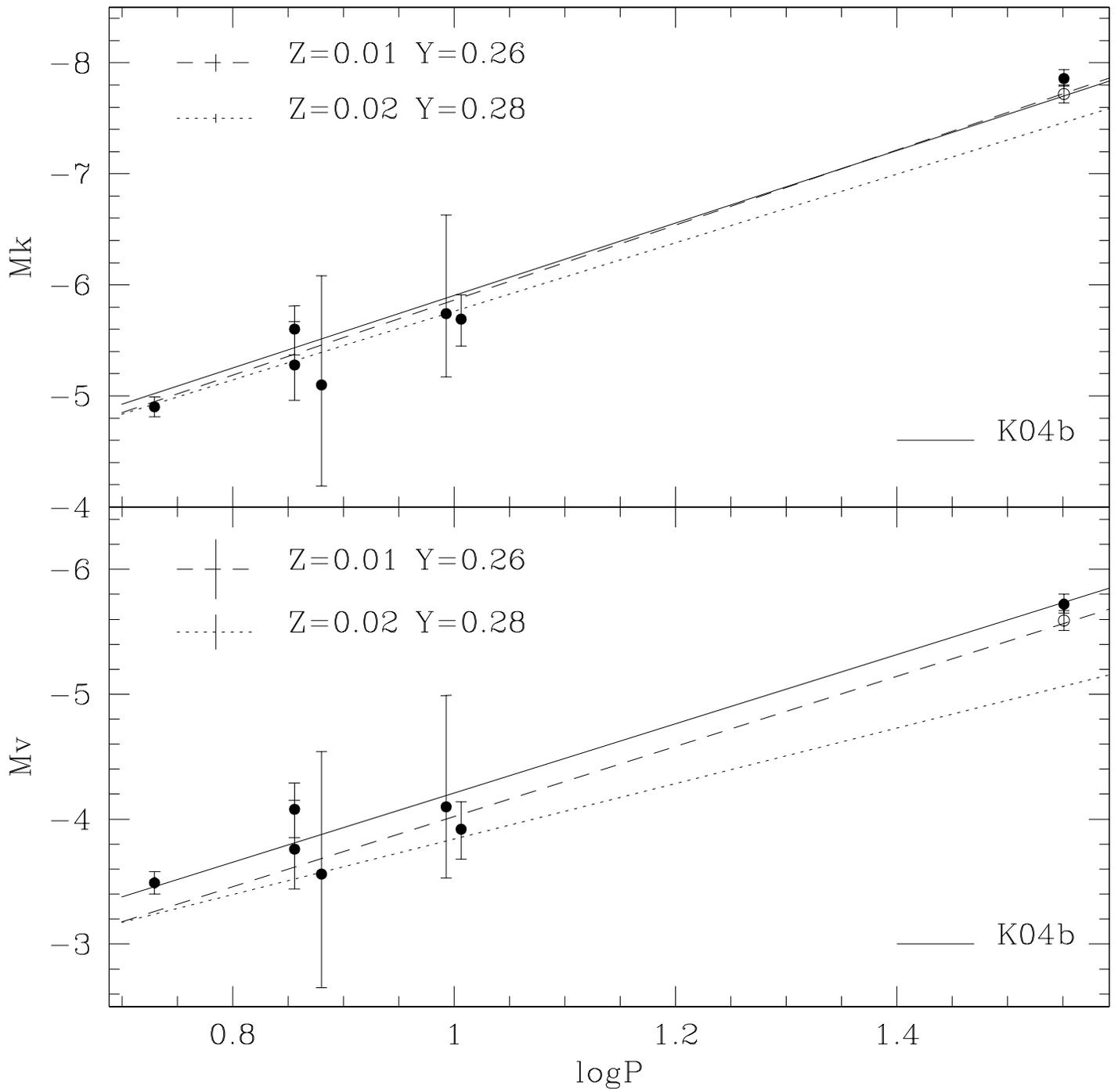
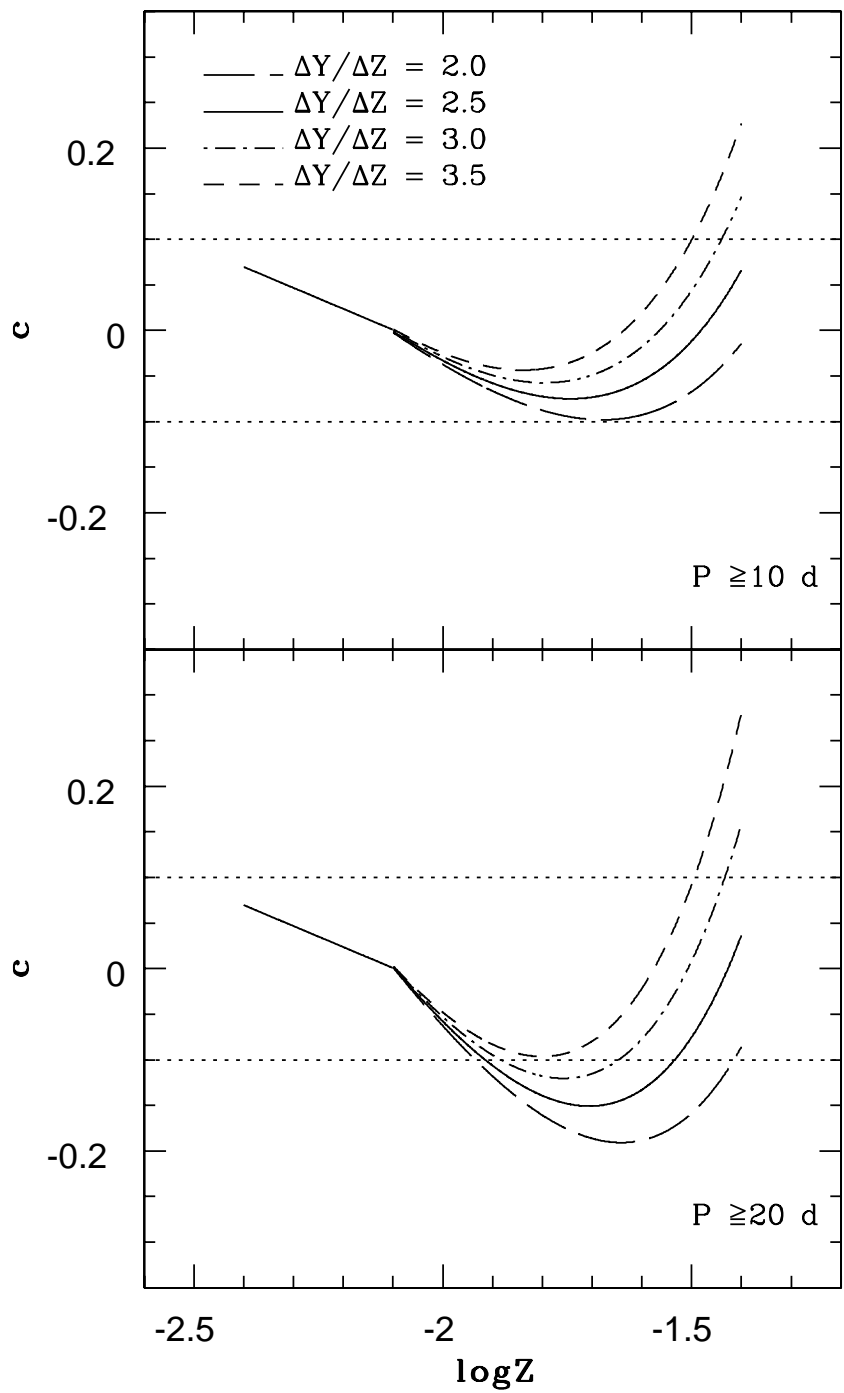
\end{document}